\documentstyle[epsf,psfig,preprint,aps]{revtex}

\ifx\undefined\psfig\def\psfig#1{ }\else\fi

\begin{document}
\ifpreprintsty\else
\twocolumn[\hsize\textwidth%
\columnwidth\hsize\csname@twocolumnfalse\endcsname
\fi

\draft
\preprint{IPM-97-261}
\title{ Diamond-Like Carbon film from Liquid Gas on Metallic Substrates
             }
\author{M.A. Vesaghi${}^{a}$
\footnote{e-mail address: vesaghi@physic.sharif.ac.ir}  and
  A. Shafiekhani${}^b$\footnote{e-mail address: ashafie@theory.ipm.ac.ir}
}
\address{        ${^a}$   Dept. of Physics,
                 Sharif University of Technology,\\
                 P.O.Box: 9161,
                 Tehran 11365, Iran\\
 ${^b}$Institute for Studies in Theoretical Physics and Mathematics\\
             P.O.Box: 5531, Tehran 19395, Iran
}

\maketitle
\begin{abstract}
\leftskip 2cm
\rightskip 2cm

Liquid gas was used to produce DLC films on Cu, Al and steel substrates by
DC plasma technique. The absorption in IR reflectance indicates, grown films
are DLC. By deconvolution of room temperature UV-visible spectra of the films
grown at 50 mtorr and 200${}^\circ$C, in addition to the spectra lines
reported for CL, PL, PLC and ESR, some new spectra lines were obtained. We
also, have seen exciton absorption line at room temperature.
\end{abstract}

\pacs{\leftskip 2cm PACS number: 71.35.C, 78.40, 81.05.T, 78.66, 
61.72.J}

\ifpreprintsty\else\vskip1pc]\fi
\narrowtext

In the last two decade the interest on diamond-like carbon (DLC) and
hydrogenated amorphous carbon (a-C:H) films have grown enormously due to
their particular and useful
 properties such as wide band gap, high thermal
conductivity, high hardness and various preparation techniques available.
However, there are some technical difficulties in
 preparation and characterization.

Various physical and chemical techniques such as, sputtering\cite{1}, 
pulsed
laser\cite{2}, radio frequency (rf)\cite{3}, electrolysis\cite{4}, microwave
enhanced\cite{5,6}, plasma beam\cite{7}, hot filament\cite{8}, direct current
(dc) discharge plasma\cite{9}, were used to make DLC and a-C:H films. The
gases which are used in most of these deposition methods are ${\rm C}_2{\rm H%
}_2$, ${\rm C}{\rm H}_4$ and ${\rm C}_2{\rm H}_6$ pure or mixed with
hydrogen. In all these techniques neutral or ionized atoms of carbon or
various hydrocarbon precursor ions have been produced. These species with
free (dangling) hand reach substrate, and those with sufficient time and
energy stick to substrate and to each other.

It is now well understood that films deposited by these techniques are
amorphous or diamond-like carbon with no-long range order in structure.
Because of the non-equilibrium nature of the growth, there must be some
vacancies in these films\cite{5}. These vacancies are produced by the impact
of the ions reaching the growth surfaces with sufficient energy, sputtering
off weak bonded species or damaging these surfaces. Also, there are
vacancies in the domain-walls of the microcrystals. The vacancies at the
surface of substrate with film are less important since the area of this
surface is much smaller than total area of domain-walls. The quality of the
film is dependent on the nature and the numbers of these defects. If one
understands the nature of such defects and is able to control their numbers,
one would be able to control the film quality and to use them as a source
for coherent luminescence and other purposes.

Cathodoluminescence (CL)\cite{10,11}, photoluminescence (PL)\cite{2,6},
positron-life time spectroscopy (PLS)\cite{12} and electron-spin resonance
(ESR)\cite{13} methods are used for vacancy identification.

The aim of this work is to show the possibility of making DLC films from
liquid gas on metallic substrates and to obtain information about vacancies
from UV-visible reflectance of DLC films.

Carbon films were deposited on polished $2\times 3 {\rm cm}$ copper, steel
and aluminium substrates from liquid gas (60\% Butane and 40\% Propane) by dc
plasma technique. The substrate temperature was 200${}^\circ$C and it was
grounded. The chamber was flashed from 1 mtorr three times prior to
deposition. The pressure and the gas flow were kept constant during
deposition. Saturation voltage was applied to anode. Double beam
Perkin-Elmer UV-visible spectrometer was used to measure the reflection and absorption
of the films. The angle of incident beam was $6^\circ$ and measurements were
done at room temperature. IR spectra was obtained at room temperature by
Perkin-Elmer IR spectrometer.

\begin{figure}
\centerline{\psfig{figure=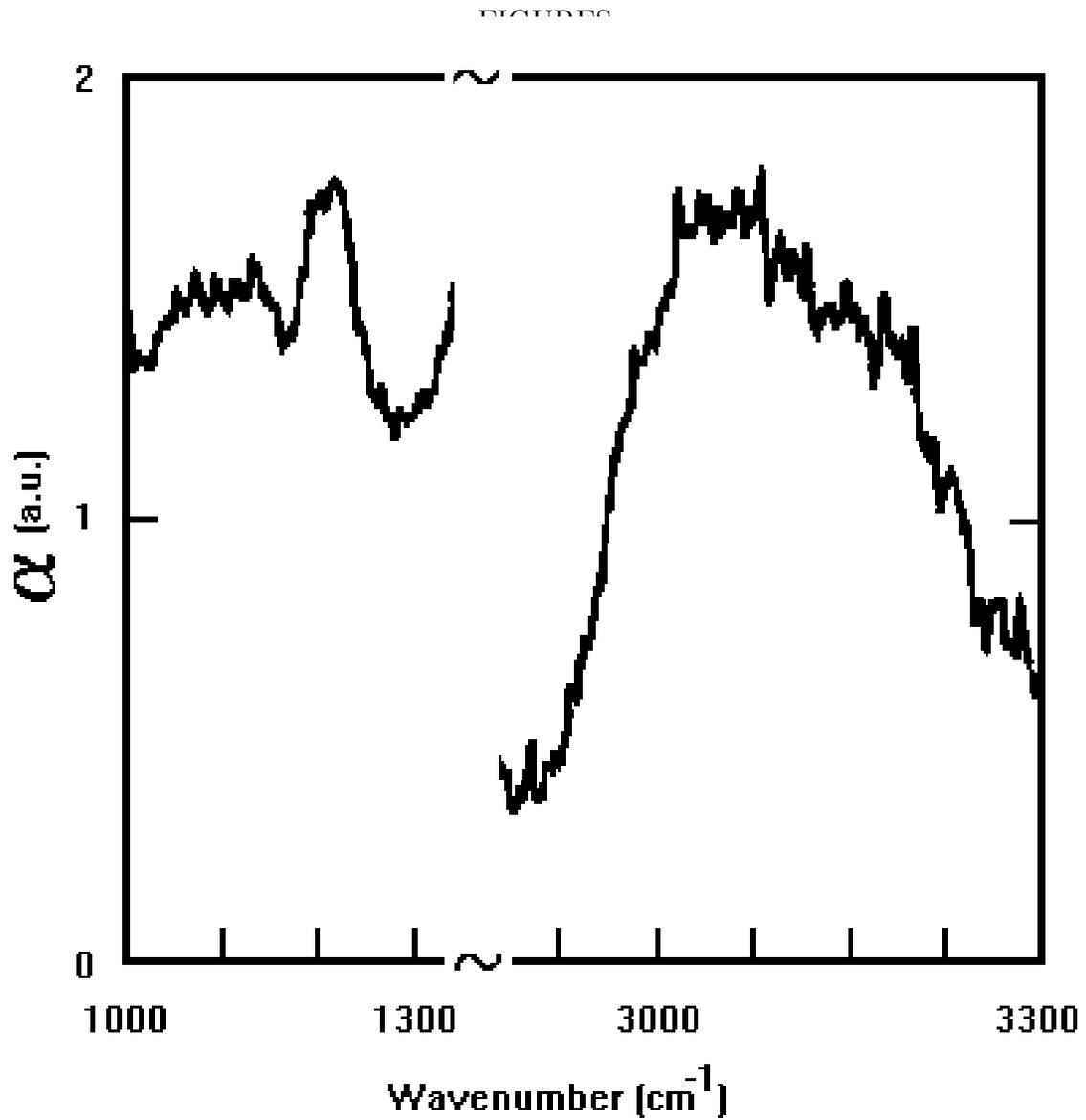,width=.9\columnwidth}}
\vspace{5mm}
\caption{Room temperature IR spectra in a reflective mode for DLC film
deposited on steel at 30 mtorr and room temperature for 1 hour.}
\end{figure}
Figure (1) shows IR reflection spectra of sample no. st8. This was DLC film
deposited on steel substrate at 30 mtorr. Most C-H vibrational modes
reported for 1000 to 1300 and 2800 to 3300 ${\rm cm}^{-1}$ by 
others\cite{3,5,7} are apparent in this figure. This film and those which 
their IR
results reported were made by various method from different hydrocarbon
gases at pressures below or above 50 mtorr. The ratio of 
$[sp^3]/[sp^2]$ for films deposited around this pressure is 
minimum\cite{9}.

\begin{figure}
\centerline{\psfig{figure=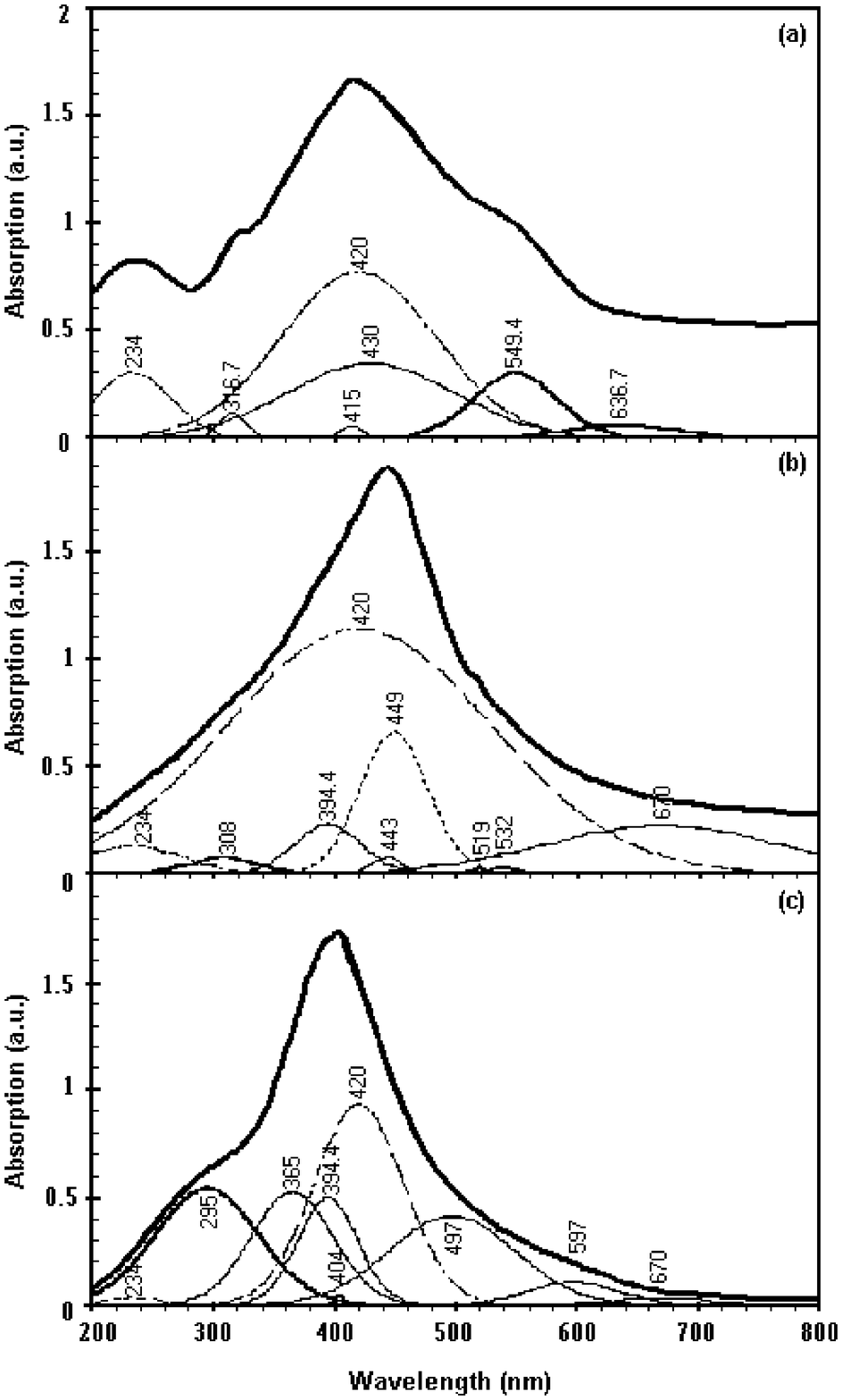,width=0.9\columnwidth}}
\vspace{5mm}
\caption{Room temperature UV-visible spectra of films deposited
at 50 mtorr and 200${}^\circ$C for 30 minutes
on (a) copper, (b) aluminium and
(c) steel. Number written on the peak of each deconvoluted normalized gaussian
is the wavelength of its maximum. The thick lines are 295, 308, 404, 549 and
637 nm lines which were not reported before plus the experimental results.
}
\end{figure}
To obtain information about vacancies in both formations, diamond and
graphite, different samples were grown at 50 mtorr. Figure (2a-c) shows
UV-visible absorption spectra of films deposited on copper, aluminium and
steel simultaneously for 30 minutes. The carbon density of these films
measured by resonance Rutherford Back-scattering (RRBS), were $2.78\times
10^{17}$, $4.06\times 10^{17}$ and $3.46\times 10^{17}$ per ${\rm cm}^2$
respectively.

The spectra of the film on Cu has two absorption peaks at 234 nm and near
420 nm. In the two other spectra, figure (2b) and (2c), only the second peak
is apparent. The peak at 234 nm (5.3 eV) is due to the recombination of free
excitons in association with momentum conserving phonons\cite{6}. This peak is
the indication of good crystallinity\cite{10,6}.
Interestingly, the exciton line is observed at room temperature.
The peak at 420 nm (2.957
ev) is due to the main vacancy. Line 420 nm exists in the spectra of nearly
all diamond-like materials produced by different methods and the natural 
diamond bombarded by high energy particles\cite{14,15,10,6}. In addition 
to these two peaks, there are other features in these figures. We fitted our
experimental graphs with the data obtained by the others from CL, PL, PLS
and ESR measurements\cite{10,11,12,13,2,6}. The deconvoluted normalized 
gaussians
are shown in the same figure. The accuracy of the fitting is better than
99.95\%. The 295, 308, 404, 549 and 637 nm lines which are needed for good
fitting were not reported before. In addition to line 420 nm which is the
strongest in all three graphs, some other lines such as 234, 430 and 549 nm
for Cu, 295, 365, 394.4 and 497 nm for Al and 449 nm for steel are
significant. Another important point is the fact that some lines are
apparent in the graphs regardless of their amplitudes and some are not. We
should emphasize that those lines which are not apparent are not the
less important ones.

Presented experimental results indicate that:\\
1- DLC films from liquid gas on metallic substrates were made.\\
2- By deconvoluting UV-visible spectra many
hidden lines have appeared and  considered for analysis of UV-visible
spectra.\\ 3- 5.3 eV absorption line of exciton at room
temperature was seen.\\

If we distinguish between the lines corresponding to the vacancies at $sp^2$
and $sp^3$ sites, we will be able to find $[sp^2]/[sp^3]$ and compare it to
the ratio obtainable from IR results or other experiments.

\vspace{0.5cm}
{\bf Acknowledgement:}

The authors are
indebted to H. Arfaei for his continuous encouragement and also,
we would like to acknowledge M. Lameie for his help on RRBS analysis.
\nopagebreak

\end{document}